\documentclass[12pt]{article}
\usepackage{graphics,cite,amssymb,epsfig,float}
\usepackage[usenames,dvips]{color}
\usepackage{citesort}
\usepackage{amsmath, mathbbol}

\newcommand\mysection{\setcounter{equation}{0}\section}

\def\ben{\begin{enumerate}}  \def\een{\end{enumerate}}

\def\lsim{\;\raise0.3ex\hbox{$<$\kern-0.75em\raise-1.1ex\hbox{$\sim$}}\;}
\def\gsim{\;\raise0.3ex\hbox{$>$\kern-0.75em\raise-1.1ex\hbox{$\sim$}}\;}

\def\ben{\begin{enumerate}}  \def\een{\end{enumerate}}
\def\bit{\begin{itemize}}    \def\eit{\end{itemize}}
\def\beq{\begin{equation}}   \def\eeq{\end{equation}}
\def\ba{\begin{array}}       \def\ea{\end{array}}
\def\bea{\begin{eqnarray}}   \def\eea{\end{eqnarray}}
\def\nn{\nonumber}

\begin{document}
\begin{titlepage}

\begin{flushright}
Report No: LPT-08-100\\
\end{flushright}

\begin{center}
\vspace{3cm}
{\Large{\bf CDF Multi-Muon Events and Singlet Extensions of the MSSM}}
\\
\vspace{1cm}

{\bf{Florian Domingo, Ulrich Ellwanger
\vspace{2cm}\\
\it 
Laboratoire de Physique Th\'eorique, CNRS -- UMR 8627,\\ 
Universit\'e de Paris--Sud, F--91405 Orsay, France}}

\begin{abstract}
We discuss a generalization of the minimal supersymmetric extension of
the Standard Model in the form of three additional singlet superfields,
which would explain the essential features of the CDF multi-muon events
presented recently: a large production cross section of $\sim 100$~pb
originates from the production of a CP-odd scalar $A$ with a mass in the
70~--~80~GeV range and a large value of $\tan\beta \sim 40$. The CP-odd
scalar $A$ decays dominantly into CP-odd and CP-even scalars $a_1$ and
$h_1$, which generate decay cascades $h_1 \to 2h_2 \to 4a_2 \to 8\tau$,
and $a_1 \to h_1 a_2$ with $h_1$ decaying as above. The decay $a_2 \to
\tau^+ \tau^-$ is slow, leading to a lifetime of ${\cal O}(20)$~ps. The
phenomenology of the model differs from similar scenarios presented
before in that one of the two cascades leads to 10 instead of 8
$\tau$-leptons, and additional production processes like associate $A$
production with $b \bar{b}$ pairs are relevant.

\end{abstract}

\vspace{2cm} \today

\end{center}
\end{titlepage}

\mysection{Introduction}

In a recent CDF publication \cite{cdf} a sample of unusual multi-muon
events was studied, which are characterized by the following properties:
i) a large rate of additional muons inside a cone of 36.8$^o$
($\cos\theta \geq 0.8$) around the direction of the trigger muon;
ii) large impact parameters (displaced vertices), and iii) an unusually
large cross section of the order of 
75~pb.\footnote{The precise value of the necessary total cross section
depends on the fraction of the events which survive the cuts applied in
\cite{cdf}, which requires a model specific simulation.}
Since neither Standard Model
processes nor known detector effects can explain the nature of these
events at present, they were refered to as ghost events.

In \cite{giromini}, some members of the CDF collaboration investigated
to which extent the properties of the ghost events can be understood in
the context of a phenomenological scenario based on cascade decays of
new particles. First, the sign-coded multiplicity distribution of
additional muons inside the $\cos\theta \geq 0.8$ cone coincides with
the assumption that originally $4 \tau^+ + 4 \tau^-$ leptons were
produced. As a hypothetical origin for the 8~$\tau$-leptons, the authors
of \cite{giromini} considered the pair production of new particles $h_1$
via $p\bar{p} \to H \to h_1 h_1$, without specifying the nature of $H$
(which could be a known or a new gauge boson, or another new particle).
Subsequently, each of the $h_1$ particles is assumed to produce the
decay cascade $h_1 \to 2 h_2 \to 4 h_3 \to 4 (\tau^+ + \tau^-)$,
generating two multi-muon cones per event. $h_2$ and $h_3$ denote
additional new states, with $h_3$ decaying as $h_3 \to \tau^+ + \tau^-$.

In order to explain the high multiplicity of additional muons {\it
inside} the
$\cos\theta \geq 0.8$ cones, the particle $h_1$ must be relatively light.
The best fit \cite{giromini} to the invariant mass distributions inside
the $\cos\theta \geq 0.8$ cones in \cite{cdf} is obtained for $h_1$,
$h_2$ and $h_3$ masses near the lower limits for which the cascade is
kinematically allowed: $m_{h_3} \sim 3.6\ \mathrm{GeV} \gsim
2\,m_{\tau}$, $m_{h_2} \sim 7.3\ \mathrm{GeV} \gsim 2\,m_{h_3}$ and
$m_{h_1} \sim 15\ \mathrm{GeV} \gsim 2\,m_{h_2}$. In order to explain
the large impact parameters, at least one of the $h_1$, $h_2$ or $h_3$
particles must have a long lifetime. The best fit \cite{giromini} to
the impact parameter distributions in \cite{cdf} corresponds to the
assumption that $h_3$ has a long lifetime of $\sim 20$~ps. The origin
of the large cross section (for the production of $H$ in this scenario)
was left unexplained in \cite{giromini}.

Some comments on the multi-muon study by the CDF collaboration
\cite{cdf} and the phenomenological interpretation in \cite{giromini}
were published in \cite{strassler}, wherein proposals for additional
studies/plots were made, and some difficulties with the phenomenological
scenario in \cite{giromini} were pointed out, on which we will comment
later. Furthermore, different phenomenological scenarios (various types
of micro-cascades $h_i \to f\bar{f'} h_{i+1}$) were proposed in
\cite{strassler}, which could provide a better fit to the data.

Theoretical models, in which multi-lepton events at hadron colliders can
be expected, have already been considered before
\cite{str1,str2,str3,str4,str5}. One class
of such models contains a rich nearly hidden sector (a ``hidden
valley''), which interacts only weakly with the Standard Model (SM)
particles, as via a heavy $Z'$ gauge boson, or via a small kinetic
mixing between the SM and the hidden sector gauge fields
\cite{str1,str2,str3,str4,str5}. It seems
possible to explain the recently observed high-energy components of
cosmic rays as remnants of dark matter annihilation in such models
\cite{nima}. (See \cite{quev} for a discussion of string vacua including
D-branes, where a light hyperweak gauge boson can connect the SM
with a hidden sector, and ghost-like events can occur.)

Multi-lepton events at hadron colliders are also possible within the
Next-to-Minimal Supersymmetric Standard Model (NMSSM) 
\cite{nmssm1,nmssm2,nmssm3,nmssm4,nmssm5,nmssm6,nmssm7},
where the Higgs sector of the MSSM is extended by a singlet superfield
$S$. The extended Higgs sector can contain a light CP-odd state $a_1$
with a mass below the $b \bar{b}$ threshold of $\sim 10.5$~GeV, such
that $a_1$ decays dominantly as $a_1 \to \tau^+ \tau^-$. If, in
addition, the SM-like Higgs scalar $h_{SM}$ decays
dominantly as $h_{SM} \to a_1 + a_1$, events with 4 $\tau$-leptons could
be the only signal of $h_{SM}$, rendering its discovery quite difficult
(see \cite{higtau1,higtau2,higtau3} and refs. therein). Alternatively,
two leptons can originate from a bino decay into a singlino-like
Lightest Supersymmetric Particle (LSP) \cite{kraml1,kraml2}. However, in
both cases the large number of muons observed by CDF is not achieved.

In the present paper we consider the extension of the Higgs sector of
the MSSM by several (three) singlets in order to discuss whether the
properties of the CDF ghost events could be understood in such a setup.
An extended Higgs sector was already implicitely suggested by the
authors of \cite{giromini} as a source of the states $h_1$, $h_2$ and
$h_3$, and the construction of corresponding models leading to the
desired masses in the range 3.5~--~15~GeV does not seem very difficult
at first sight. However, in practice various problems appear: First, the
unusually large production cross section must be explained. Second, the
desired decay channels must be dominant, and at least one large lifetime
must occur. Third, and most importantly, present constraints from
colliders (notably LEP), B~physics etc. on such an extended Higgs sector
must be satisfied. In the worst case it may be impossible to satisfy all
these conditions simultaneously, at least once the masses and couplings
of the extended Higgs sector are constrained by supersymmetry.

Hence, it is important to look for a ``go-theorem'': a concrete model,
which has all these desired properties. Here we present such a model,
which has the following structure: The starting point is the NMSSM 
involving a singlet superfield $S$, whose vacuum expectation value (vev)
solves the $\mu$-problem of the MSSM. We consider a region in the
parameter space of the NMSSM, where the lightest CP-odd Higgs state in
the $H_u$-$H_d$-$S$ sector (denoted by $A$ subsequently) has a mass in
the 70~--~80~GeV range and has both large singlet and large doublet
components of $\sim 85\%$ and $\sim 50\%$ respectively. Assuming a large
value of $\tan\beta \sim 40$, the production cross section of $A$ via
gluon-gluon fusion at the Tevatron can be $\sim 100$~pb
\cite{abdel2}. In
the corresponding region in parameter space, the CP-even Higgs states in
the $H_u$-$H_d$-$S$ sector as well as the second CP-odd state have
masses above 114~GeV, in which case all LEP constraints on various Higgs
production processes \cite{lep} are satisfied. (The precise Higgs masses
as well as various B-physics observables depend on the soft
supersymmetry breaking gaugino, squark and slepton masses and couplings,
for which ranges of desired values can be found without particular
effort.)

To the Higgs sector of the NMSSM we add two more singlets $S_1$ and
$S_2$, which contain two more CP-even states $h_{1,2}$ and two more
CP-odd states $a_{1,2}$. (From here onwards, the indices 1,2 denote the
additional singlets of the model rather than the states introduced in
\cite{giromini}.) These supplementary fields allow for many additional
Yukawa couplings and soft terms, but for simplicity we assume that most
of the possible Yukawa couplings vanish or are negligibly small
(considering small Yukawa couplings, as those appearing in the SM, as
natural). Then, the following situation can be achieved without fine
tuning, assuming corresponding values of the additional soft terms: The
mass matrices of the $h_{1,2}$ and $a_{1,2}$ states are nearly diagonal,
with eigenvalues in the 3.5~--~20~GeV range. A Yukawa coupling between
$S$ and $S_1$ remains relatively large, and as a consequence the CP-odd
state $A$ of the $H_u$-$H_d$-$S$ sector decays dominantly into $h_1 +
a_1$ (as compared to $A \to b \bar{b}$). Subsequently, due to a small
$S_1$-$S_2$ mixing, $h_1$ decays as $h_1 \to 2h_2 \to 4a_2 \to 8\tau$,
and $a_1$ as $a_1 \to h_1 a_2$ with $h_1$-decays as above. 

The decay $a_2 \to \tau^+ \tau^-$ is possible due to a tiny $S_{2}-H_d$
mixing of ${\cal O}(10^{-5})$, implying a $a_2$ lifetime of  ${\cal
O}(20)$~ps. Assuming $m_{a_1} \sim 20$~GeV and $m_{h_1} \sim 15$~GeV,
the first $A \to h_1 a_1$ decay will generate two separate cones
containing 8 or 10 $\tau$-leptons (which are thus not completely
symmetric) as well as displaced vertices. Thus, the essential properties
of the phenomenological scenario of \cite{giromini} are reproduced, and
the required cross section is obtained.

On the other hand the concrete model makes it clear that a
phenomenological analysis based on a single process can be incomplete or
even misleading: In the present scenario, the CP-odd state $A$ will also
be produced in association with $b \bar{b}$ pairs with a cross section
of $\sim 30\%$ of the production via gluon-gluon fusion
\cite{abdel2}. In
$\sim 25\%$ of these cases, a $b$ or $\bar{b}$ decay will generate at
least one additional muon, which can end up in one of the cones defined
by the trigger muons. These additional muons will have an impact on
observables like $\sum |p_T|$ and invariant masses, whose precise effect
can only be studied with the help of simulations, which are
beyond the scope of the present paper. (Likewise, the production and the
decays of the heavier states of the $H_u$-$H_d$-$S$ sector can
contribute to the observables.) As already underlined in
\cite{strassler}, an analysis of the data involving more stringent cuts
than the ones presented in \cite{cdf} (as muon charge selection rules)
should make it easier to constrain -- or to verify -- concrete models as
the one discussed here.

We are aware of the fact that the properties of the CDF multi-muon event
samples still need to be confirmed, notably by the D0 collaboration.
Nevertheless we found it useful to develop a concrete model, which
indicates which additional complications can be expected and which, in
any case, extends the perimeter of possible signals that may be expected
at hadron colliders. Of course, if the properties of the multi-muon
event samples are confirmed, it is also interesting to know that
relatively simple singlet extensions of the MSSM can generate such
signals. In the next section we will present the Lagrangian and discuss
the parameters of the model. Its phenomenology and conclusions will be
presented in section 3.

\mysection{A toy model}

As mentioned in the introduction, we consider a supersymmetric extension
of the SM involving an extended Higgs sector, which consists
of the MSSM doublets $H_u$, $H_d$ and three singlet superfields $S$,
$S_1$ and $S_2$. A priori, a large number of terms could appear in the
superpotential $W$, even after the restriction to scale invariant
Yukawa couplings as motivated by a solution of the $\mu$-problem of
the MSSM. On the other hand, small or vanishing Yukawa couplings are
``technically natural'' (stable under quantum corrections) in
supersymmetry, and we use this freedom to omit most of the allowed
couplings.

The relevant Higgs dependent terms in the superpotential $W$ are assumed
to be given by 
\beq\label{supot}
W = \lambda S H_u H_d + \frac{\kappa}{3} S^3 + \lambda_1 S S_1^2 +
\frac{\kappa_1}{3} S_1^3 + \lambda_2 S_1 S_2^2 + \frac{\kappa_2}{3}
S_2^3\; .
\eeq

The first two terms correspond to the ones of the NMSSM 
\cite{nmssm1,nmssm2,nmssm3,nmssm4,nmssm5,nmssm6,nmssm7},
where the vev $s \equiv \left<S\right>$ generates an effective
$\mu$-parameter $\mu_{\mathrm{eff}} = \lambda s$. The terms proportional
to $\kappa_1$ and $\kappa_2$ serve to stabilize the scalar potential for
the vevs $s_1 \equiv \left<S_1\right>$ and  $s_2 \equiv
\left<S_2\right>$, and $\lambda_1$ and $\lambda_2$ induce couplings
between $S$ and $S_1$, and between $S_1$ and $S_2$, respectively.

The soft terms in the Higgs sector are the scalar masses squared  and
trilinear couplings
\beq\label{soft}
m_{H_u}^2,\ m_{H_d}^2,\ m_S^2,\ m_{S_1}^2,\ m_{S_2}^2,\ A_\lambda,\
A_\kappa,\ A_{\lambda_1},\ A_{\kappa_1},\ A_{\lambda_2}\ \mathrm{and}\
A_{\kappa_2}\; .
\eeq

It is straightforward to work out the tree level Higgs mass matrices and
couplings from the superpotential and the soft terms, which leads to
quite lengthy and not very transparent expressions. Instead of
presenting them, we first consider the ``decoupling limit'' $\lambda_2
\to 0$. As it becomes evident from the superpotential, all components of
the superfield $S_2$ decouple from $H_u$, $H_d$, $S$ and $S_1$ in this
limit. Furthermore one finds that for a wide range of parameters (for
$A_{\kappa_1}^2$ not too large and positive $m_{h_1}^2$, see
eq.~(\ref{masses}) below), the vev $s_1$ vanishes, since terms linear in
$s_1$ in the scalar potential are proportional to $\lambda_2$. Then the
mass matrices in the Higgs sector are block-diagonal: 

In the NMSSM sector $H_u$-$H_d$-$S$ one re-obtains the well known $3
\times 3$ ($2 \times 2$) mass matrices for the CP-even (CP-odd) states
\cite{nmssm1,nmssm2,nmssm3,nmssm4,nmssm5,nmssm6,nmssm7}. 
In addition, the mass matrices for the CP-even $h_{1,2}$
and CP-odd $a_{1,2}$ states are diagonal. In the case of the vev
$s_2$, we assume that $|A_{\kappa_2}|$ is sufficiently large such that
the vev $s_2$ is nonvanishing (which avoids degenerate $h_2$, $a_2$
states and a massless neutralino $\psi_2$). Then it is convenient to
express $m_{S_2}^2$ in terms of $s_2$,
$\kappa_2$ and $A_{\kappa_2}$ through the minimization equation
of the scalar potential, after which the masses of the physical
states $h_{1,2}$ and $a_{1,2}$ can be written as
\bea\label{masses}
m_{h_1 }^2 &=& m_{S_1}^2 + 2 \lambda_1 A_{\lambda_1} s + 2 \kappa
\lambda_1 s^2 -2 \lambda \lambda_1 v_u v_d + 4 \lambda_1^2 s^2\; ,\nn \\
m_{a_1 }^2 &=& m_{S_1}^2 - 2 \lambda_1 A_{\lambda_1} s - 2 \kappa
\lambda_1 s^2 +2 \lambda \lambda_1 v_u v_d + 4 \lambda_1^2 s^2\; ,\nn \\
m_{h_2 }^2 &=& \kappa_2 s_2 (A_{\kappa_2} + 4 \kappa_2 s_2)\; ,\nn \\
m_{a_2 }^2 &=& -3 \kappa_2 A_{\kappa_2} s_2\
\eea
where $v_u$, $v_d$ denote the vevs of the neutral components of $H_u$,
$H_d$.

Evidently there exist sufficient free parameters $m_{S_1}^2$,
$\lambda_1$, $A_{\lambda_1}$, $s_2$, $\kappa_2$ and $A_{\kappa_2}$ in
the $S_1$-$S_2$ sector in order to generate a spectrum like
\beq
 m_{a_1} \sim 20\ \mathrm{GeV}\;,\ m_{h_1} \sim 16\ \mathrm{GeV}\;,\ 
 m_{h_2} \sim 8\ \mathrm{GeV}\;, \ m_{a_2} \sim 4\ 
 \mathrm{GeV}\;,
 \eeq
which render the cascade decays described in
the introduction kinematically possible, with relatively light initial
states $h_1$ and $a_1$. (The masses of the additional neutralinos are
given by $m_{\psi_1} = 2 \lambda_1 s$ and $m_{\psi_2} = 2 \kappa_2 s_2$.
$\psi_1$ is too heavy to be produced in $A$~decays, and the ${\rm BR}(A
\to \psi_2\, \psi_2)$ vanishes in the decoupling limit.)

Of course, the desired cascade decays of $h_1$ and $a_1$ require the
presence of couplings $g_{h_1 h_2 h_2}$, $g_{a_1 h_1 a_2}$ and $g_{a_2
\tau^+ \tau^-}$, which are absent in the decoupling limit $\lambda_2 \to
0$. (The coupling $g_{h_2 a_2 a_2}$ is of the order $\kappa_2
A_{\kappa_2}$ and not suppressed in the decoupling limit.) One can check
that a small value of $\lambda_2$ will generate couplings of the order
(modulo Yukawa couplings and a dimensionful parameter like an $A$-term
or a vev) $g_{h_1 h_2 h_2} \sim g_{a_1 h_1 a_2} \sim \lambda_2$, $g_{a_2
\tau^+ \tau^-} \sim \lambda_2^2$. One finds that for $\lambda_2^2
\sim 10^{-5}$, the $a_2$ lifetime will be of the order 20~ps as desired,
but which has no noticeable effect on the eigenvalues of the mass
matrices above.

In the NMSSM sector, the tree level mass matrices receive considerable
radiative corrections depending on the squark, slepton and gaugino
masses and trilinear couplings. These are included in the code
NMHDECAY/NMSSMTools \cite{nmhdecay1,nmhdecay2,nmhdecay3}, which computes the NMSSM Higgs
masses and couplings as functions of the parameters in the Lagrangian.
As independent parameters in the Higgs sector of the NMSSM, one can
choose \cite{nmhdecay1,nmhdecay2,nmhdecay3}
\beq\label{paras}
\lambda,\ \kappa,\ A_\kappa,\ \tan\beta,\ \mu_{\mathrm{eff}} \equiv
\lambda s,\ M_A^2 \equiv \frac{2 \mu_{\mathrm{eff}}(A_\lambda + \kappa
s)}{\sin 2\beta}\; .
\eeq

Large cross sections of Higgs particles at hadron colliders occur at
large values of $\tan\beta$, for which the coupling of $H_d$ to down
quarks is proportional to $\tan\beta$. Then, the $b$~quark loop induced
gluon-gluon fusion process has a cross section amplified by  $\sim
\tan^2\beta$ with respect to the corresponding cross section for the
production of a SM Higgs scalar. On the other hand, the
cross section via gluon-gluon fusion decreases strongly with increasing
Higgs masses, but low CP-even Higgs masses are strongly constrained by
LEP \cite{lep}. Therefore we concentrate on a region in the NMSSM
parameter space (\ref{paras}) at large $\tan\beta$ where a CP-odd Higgs
scalar $A$ has a mass $m_A$ below 100~GeV, but large enough to render
the decay $A \to h_1 a_1$ (with $A$ on-shell) possible. Also, the decay
$A \to h_1 a_1$ should have a larger branching ratio than the decay $A
\to b \bar{b}$, which requires a considerable singlet $S$ component of
$A$ without a too large suppression of the coupling of $A$ to $b$~quarks
(see the next section).

Finally, LEP constraints on all CP-even Higgs scalars as well as
constraints from B~physics (the branching ratios ${\rm BR}(B\to
X_s\gamma)$, ${\rm BR}(\bar{B}^+ \to \tau^+\nu_{\tau})$, ${\rm BR}(B_s
\to \mu^+\mu^-)$ and the mass differences $\Delta M_q$, $q=d,s$), which
are particularly relevant at large $\tan\beta$, should be satisfied. All
these constraints are checked in the code NMHDECAY/NMSSMTools
\cite{nmhdecay1,nmhdecay2,nmhdecay3}, which we used in the
search for acceptable regions in the parameter space (\ref{paras}) of
the model. Clearly, Higgs masses as well as B~physics observables depend
also on the soft supersymmetry breaking gaugino, squark and slepton
masses and couplings, which have to be specified.

In fact, for large $\tan\beta \sim 40$ and non-negligible NMSSM Yukawa
couplings $\lambda$ and $\kappa$ a phenomenologically acceptable region
in the parameter space exists, in which $A$ has a mass in the
70~--~80~GeV range and $H_d$- and $S$-components of $\sim 50\%$ and
$\sim 85\%$, respectively. An example is given by the following point in
parameter space, where
\beq\label{paraspt}
\lambda=.28,\ \kappa=.33,\ A_\kappa=-36.5,\ \tan\beta=40,\
\mu_{\mathrm{eff}} = 240\,\mathrm{GeV},\ M_A = 420\,\mathrm{GeV}\; .
\eeq
The gaugino masses are $M_1=150$~GeV, $M_2=300$~GeV, $M_3=1$~TeV,
the left-handed and right-handed up-type squark masses are 1.5~TeV, the
right-handed down-type squark masses are given by 1~TeV, the slepton
masses by 500~GeV, $A_{top} = A_{bottom} = 1.8$~TeV and $A_\tau =
300$~GeV.

For these parameters, the lightest CP-odd Higgs mass $m_A$ and its
decomposition $A =  N_{A,A_u} A_u + N_{A,A_d} A_d + N_{A,S} A_s$ (where
$A_u$, $A_d$ and $A_s$ are the neutral CP-odd components of $H_u$, $H_d$
and $S$) are given by
\beq\label{aprop} m_A = 70\,\mathrm{GeV},\ N_{A,A_u} = 0.01,\  N_{A,A_d}
= 0.56,\ N_{A,S} = 0.83\; .
 \eeq

The masses of the three CP-even Higgs scalars are 114.5~GeV, 270~GeV and
561~GeV, and the masses of the second CP-odd and charged Higgs scalars
are 300~GeV and 264~GeV, respectively. Further properties of this point
in parameter space, which are relevant for the CDF ghost events, will be
discussed in the next section.


\mysection{Phenomenology of the toy model}

In order to estimate the production cross section of $A$ via gluon-gluon
fusion at the Tevatron, one has to determine its reduced coupling $X_d$
to down-type quarks (normalized to the SM Higgs coupling),
\beq\label{xd}
X_d = \tan\beta\times N_{A,A_d}\qquad ( = 22.2)\, ,
\eeq
where the value in parenthesis is the one for the point given in
(\ref{paraspt}), (\ref{aprop}). Then, the corresponding production cross
sections for the Tevatron in \cite{abdel2} can be rescaled appropriately,
and extrapolated to $m_A =$ 70~--~80~GeV. For $m_A = 70$~GeV and $X_d
\sim 22$ one obtains $\sigma (p\bar{p} \to A+X) \sim 100$~pb, even
somewhat larger than required.

Subsequently, we have to estimate the $A$ decay branching fractions. In
the absence of the $S_{1,2}$ sector, $A$ would decay to $\sim 90\%$ into
$b \bar{b}$ with a partial width
\beq
\Gamma_{b \bar{b}} = \frac{3\, G_F}{4\, \sqrt{2}\, \pi} X_d^2\, m_b^2\,
m_A\, \sqrt{1-\frac{4\, m_b^2}{m_A^2}}\; ,
\eeq
which gets enhanced by $\sim 20\%$ by QCD corrections. In the presence
of a coupling $g_{A h_1 a_1}$, the partial width for the decay $A \to
h_1 a_1$ is
\beq
\Gamma_{h_1 a_1} = \frac{g_{A h_1 a_1}^2}{16\,\pi\,m_A}\,
\sqrt{\left(1-\frac{m_{h_1}^2}{m_A^2}-\frac{m_{a_1}^2}{m_A^2}\right)^2
-4\frac{m_{h_1}^2 m_{a_1}^2}{m_A^4}}\; .
\eeq
Numerically, one obtains for the ratio
\beq
R = \frac{\Gamma_{h_1 a_1}}{\Gamma_{b \bar{b}}} \sim \left(\frac{36\, 
g_{A h_1 a_1}}{X_d\, m_A}\right)^2\; .
\eeq
In order to obtain a branching fraction for $A\to h_1 a_1$ larger than
$\sim 80\%$, such that the production cross section $\sigma (p\bar{p}
\to A\to h_1 a_1)$ is larger than $\sim 80$~pb, we should have $R\gsim
4$.\footnote{We recall the footnote on page~1, according to which the
necessary total cross section can be somewhat larger or smaller.}
For the values of $X_d$ and $m_A$ above, this can be obtained for
$g_{A h_1 a_1} \gsim 86$~GeV. In the present model, $g_{A h_1 a_1}$ is
given by
\beq
g_{A h_1 a_1} = \frac{N_{A,S}}{\sqrt{2}}\lambda_1 \left(A_{\lambda_1} +
2 \kappa s\right)
\eeq
with $N_{A,S}$ as in (\ref{aprop}). Note that $g_{A h_1 a_1}$ is not
suppressed in the decoupling limit $\lambda_2 \to 0$. It is easy to find
values for $\lambda_1 \sim 0.52$ and $A_{\lambda_1} \sim -283$~GeV such
that $g_{A h_1 a_1}$ is sufficiently large, and the masses $m_{h_1}$ and
$m_{a_1}$ obtained from (\ref{masses}) have the desired values.

After the dominant decay $A\to h_1 a_1$, $h_1$ and $a_1$ cannot decay
at tree level in the decoupling limit $\{\lambda_2,\ s_1\} \to 0$, where
$h_1$ and $a_1$ do not mix with the $H_u$-$H_d$-$S$ sector. Small values
of $\lambda_2$ (and appropriate natural values for $A_{\lambda_2},\
\kappa_2$ and $A_{\kappa_2}$) are sufficient in order to generate the
dominant decays $h_1 \to h_2 h_2$, $h_2 \to a_2 a_2$ and $a_1 \to h_1
a_2$, which we
assume to be kinematically allowed, and which produce the cascades
described in the introduction. The decay of $a_2$ into quarks and
leptons is made possible only through its small mixing $\sim
\lambda_2^2$ with the $H_u$-$H_d$-$S$ sector. For large $\tan\beta$,
$a_2$ mixes dominantly with $H_d$, from which it inherits the couplings
proportional to the down-type fermion masses leading to the dominant
decay $a_2 \to \tau^+ \tau^-$ (for $2\, m_b > m_{a_2} \gsim 2\,
m_\tau$). Herewith we have reproduced the essential features of the
scenario proposed in \cite{giromini}, albeit with one of the two 
cascades (the one originating from $a_1$) leading to 10 rather than 8
$\tau$-leptons in the final state, which will evidently imply some
modifications of the plots presented in \cite{giromini}.

Next we comment on an issue raised in \cite{strassler}, where it has
been noted that a Higgs-like coupling to $\tau$-leptons implies a
coupling to muons with a ratio $m_\mu/m_\tau$, and hence a ratio of
branching ratios ${\rm BR}(a_2 \to \mu^+ \mu^-)/{\rm BR}(a_2 \to \tau^+
\tau^-) \gsim m_\mu^2/m_\tau^2 \sim 0.0035$. For $m_{a_2}$ near $2\,
m_\tau$ the ratio of branching ratios increases due to the kinematic
suppression of the ${\rm BR}(a_2 \to \tau^+ \tau^-)$. In plots of the
invariant
mass of muons of opposite charges inside a cone, this could (should)
generate a visible peak at $m_{a_2}$. This reasoning remains valid in
our case, but we note that the kinematic suppression of the ${\rm
BR}(a_2 \to \tau^+ \tau^-)$ near the threshold $m_{a_2} \to 2\, m_\tau$
is less important for CP-odd scalars (like $a_2$) as compared to CP-even
scalars; accordingly, the ratio of branching ratios above will increase
less dramatically near the threshold $m_{a_2} \to 2\, m_\tau$, which
makes it somewhat more difficult to rule out the scenario through a
non-observation of a peak in the $\mu^+ \mu^-$ invariant mass
distribution.

Finally we turn to the invariant mass distribution $M$ of all muons --
or of all tracks -- for events in which both cones contain at least two
muons (Fig.~35 in \cite{cdf}). According to the simulations of the
process $p\bar{p} \to H \to h_1h_1$ performed in \cite{giromini} (in the
notation of \cite{giromini}), a resonance-like structure should be
visible with a peak position depending on $m_H$ (see Fig.~6 in
\cite{giromini}); however, the data do not show such a structure: the
process simulated in  \cite{giromini} could not describe simultaneously
the steep rise of the invariant mass distribution for small $M$, and the
tail of the invariant mass distribution at large $M$, for any value of
$m_H$.

In the present scenario we have to replace $H$ by $A$ with a mass in the
70~--~80~GeV range, which seems to describe only the invariant mass
distribution for small $M$. Furthermore, one of the cones would
contain 10 $\tau$-leptons; however, this is possibly not yet enough in
order to explain the tail of the invariant mass distribution at large
$M$. 

On the other hand, as already mentioned in the introduction, there
exist additional production processes which have necessarily to be taken
into account: the cross section for associate $b\bar{b} + A$ production
can be estimated to be $\sim 30\%$ of the $A$ production via gluon-gluon
fusion (hence $\sim 30$~pb) for $A$ masses in the range considered here
\cite{abdel2}. In $\sim 25\%$ of these cases, a $b$ or
$\bar{b}$ decay will generate at least one additional muon which can
contribute to the tail of the invariant mass distribution at large $M$.

Furthermore, one of the heavier CP-even scalars (the one with a mass of
$\sim 270$~GeV for the point above) as well as the heavier CP-odd scalar
$A_{heavy}$ (with a mass $\sim 300$~GeV here) have couplings to
$b$~quarks enhanced by factors of $\tan\beta = 40$ and $\tan\beta \times
N_{A_{heavy},A_d} \sim 33$, respectively. At least in regions in
parameter space where their masses are still lower, these states --
which will generate similar cascades leading to 8 -- 10 $\tau$-leptons
-- can also contribute to the tail of the invariant mass distribution at
large $M$. Of course, further simulations are necessary in order to
check these conjectures, but at first sight explanations of the
invariant mass distributions for small and for large $M$ seem possible.
Clearly, these processes will also contribute to other observables like
$\sum p_T$.

At last, a comment on the dark matter relic density in this model is
appropriate. The LSP is the neutralino $\psi_2$ with a mass $m_{\psi_2}
= 2 \kappa_2 s_2 \approx m_{h_2} \gsim 2 m_{a_2}$. Its annihilation is
dominated by the processes $\psi_2 + \psi_2 \to h_2 \to a_2 + a_2$ and
$\psi_2 + \psi_2 \to a_2 \to h_2 + a_2$. Subsequently the scalars
$a_2$ ($h_2$) will decay into two (four) $\tau$-leptons as at the end of
the cascades relevant for the multi-muon events. The annihilation
processes depend on the $\psi_2\,\psi_2\,h_2/a_2$ Yukawa coupling
$\kappa_2$, and on the trilinear coupling $g_{h_2 a_2 a_2}$ of the order
of $\kappa_2 A_{\kappa_2}$. Whereas $A_{\kappa_2}$ is
determined by the desired values of $m_{h_2}$ and $m_{a_2}$
(\ref{masses}) to be $A_{\kappa_2} \sim - (1 - 1.5)$~GeV (and
$\kappa_2 s_2 \sim 4$~GeV), the value of $\kappa_2$ is unconstrained so
far. One can expect that, for a value of $\kappa_2$ in the range ${\cal
O}(10^{-3})$ -- ${\cal O}(10^{-1})$, the WMAP value $0.094 \, \lesssim
\Omega_{\psi_2} h^2 \lesssim 0.136$ \cite{wmap} for the dark matter
relic density can be achieved.

To conclude, apart from the fact that the CDF multi-muon events
\cite{cdf} need to be confirmed notably by the D0 collaboration,
additional studies of their properties would be desirable, as the ones
pointed out in \cite{strassler}: spatial correlations among displaced
vertices, and invariant mass distributions of dimuon pairs depending on
their relative charges. In any case, plots have to be compared with
simulations of models. 

We have presented a relatively simple model in the form of a
multi-singlet extension of the MSSM, whose particle content and
parameters have been chosen such that the essential features of the CDF
multi-muon events  can be reproduced, without contradicting constraints
from other experiments. Already in this scenario, the phenomenology
would be more complicated than the one discussed in \cite{giromini}.
Clearly, the particular values of the parameters of the model chosen in
section~2 -- and the corresponding masses and couplings -- have been
presented for illustrative purposes only, and eventually the complete
phenomenologically acceptable region in parameter space could be
studied. Together with further models, which will certainly be proposed
soon, this will allow for comparisons or ``best fits'' to the data.

\section*{Acknowledgements}

It is a pleasure to thank A. Djouadi and A.M.~Teixeira for helpful
discussions.



\begin{thebibliography}{99}

\bibitem{cdf} T.~Aaltonen {\it et al.}  [CDF Collaboration], ``Study of
  multi-muon events produced in p-pbar collisions at sqrt(s)=1.96 TeV,''
  arXiv:0810.5357 [hep-ex].

\bibitem{giromini} P.~Giromini, F.~Happacher, M.~J.~Kim, M.~Kruse,
  K.~Pitts, F.~Ptohos and S.~Torre, ``Phenomenological interpretation of
  the multi-muon events reported by the CDF collaboration,''
  arXiv:0810.5730 [hep-ph].

\bibitem{strassler} M.~J.~Strassler, ``Flesh and Blood, or Merely
  Ghosts? Some Comments on the Multi-Muon Study at CDF,''
  arXiv:0811.1560 [hep-ph].
  
\bibitem{str1} 
    M.~J.~Strassler and K.~M.~Zurek,
  Phys.\ Lett.\  B {\bf 651} (2007) 374
  [arXiv:hep-ph/0604261].
  
\bibitem{str2} 
    M.~J.~Strassler and K.~M.~Zurek,
  Phys.\ Lett.\  B {\bf 661} (2008) 263
  [arXiv:hep-ph/0605193].
  
\bibitem{str3} 
  M.~J.~Strassler, ``Possible effects of a hidden valley on
  supersymmetric phenomenology,'' arXiv:hep-ph/0607160.
  
\bibitem{str4} 
  T.~Han, Z.~Si, K.~M.~Zurek and M.~J.~Strassler,
  JHEP {\bf 0807} (2008) 008
  [arXiv:0712.2041 [hep-ph]].
  
\bibitem{str5} 
N.~Arkani-Hamed and N.~Weiner,
  JHEP {\bf 0812} (2008) 104
  [arXiv:0810.0714 [hep-ph]].
  
\bibitem{nima}
  N.~Arkani-Hamed, D.~P.~Finkbeiner, T.~Slatyer and N.~Weiner,
  ``A Theory of Dark Matter,''
  arXiv:0810.0713 [hep-ph].
  
\bibitem{quev}
 J.~P.~Conlon, A.~Maharana and F.~Quevedo,
  ``Towards Realistic String Vacua,''
  arXiv:0810.5660 [hep-th].
  
\bibitem{nmssm1} 
P. Fayet, Nucl. Phys. B \textbf{90} (1975) 104.

\bibitem{nmssm2} 
P. Fayet, 
Phys. Lett. B \textbf{69} (1977) 489.

\bibitem{nmssm3} 
H.P. Nilles, M. Srednicki and D. Wyler, 
Phys. Lett. B \textbf{120} (1983) 346.

\bibitem{nmssm4} 
J.M. Frere, D.R. Jones and S. Raby, 
Nucl. Phys. B \textbf{222} (1983) 11.

\bibitem{nmssm5}  
J.P. Derendinger and C.A. Savoy, Nucl. Phys. B \textbf{237} (1984) 307.

\bibitem{nmssm6}  
J. Ellis, J. Gunion, H. Haber, L. Roszkowski and F. Zwirner, 
Phys. Rev. D \textbf{39}  (1989) 844.

\bibitem{nmssm7} 
M. Drees, Int. J. Mod. Phys. A \textbf{4}  (1989) 3635.

\bibitem{higtau1} 
A.~Djouadi {\it et al.},  JHEP {\bf 0807} (2008) 002
  [arXiv:0801.4321 [hep-ph]].
  
\bibitem{higtau2} 
  S.~Chang, R.~Dermisek, J.~F.~Gunion and N.~Weiner,
   Ann.\ Rev.\ Nucl.\ Part.\ Sci.\  {\bf 58} (2008) 75
  [arXiv:0801.4554 [hep-ph]].
  
\bibitem{higtau3} 
A.~Belyaev {\it et al.}, ``The Scope of the 4 tau Channel in
  Higgs-strahlung and Vector Boson Fusion for the NMSSM No-Lose Theorem
  at the LHC,'' arXiv:0805.3505 [hep-ph].

\bibitem{kraml1}  K.~Cheung and T.~J.~Hou,
  ``Light Pseudoscalar Higgs boson in Neutralino Decays in the
  Next-to-Minimal Supersymmetric Standard Model,''
  arXiv:0809.1122 [hep-ph].
  
\bibitem{kraml2}
  S.~Kraml, A.~R.~Raklev and M.~J.~White, ``NMSSM in
  disguise: discovering singlino dark matter with soft leptons at the
  LHC,'' arXiv:0811.0011 [hep-ph].

\bibitem{abdel2}
A.~Djouadi,
  Phys.\ Rept.\  {\bf 459} (2008) 1 [arXiv:hep-ph/0503173].
  
\bibitem{lep} 
S.~Schael {\it et al.}  [ALEPH Collaboration and DELPHI Collaboration
and L3 Collaboration and OPAL Collaboration],
Eur.\ Phys.\ J.\  C {\bf 47} (2006) 547 [arXiv:hep-ex/0602042].

\bibitem{nmhdecay1}  U.~Ellwanger, J.~F.~Gunion and C.~Hugonie,  JHEP
{\bf 0502} (2005) 066; see
the web site {\sf http://www.th.u-psud.fr/NMHDECAY/nmssmtools.html}.


\bibitem{nmhdecay2}
  U.~Ellwanger and C.~Hugonie,  Comput.\ Phys.\
Commun.\  {\bf 175} (2006) 290 [arXiv:hep-ph/0508022].

\bibitem{nmhdecay3}
  U.~Ellwanger and C.~Hugonie,  Comput.\
Phys.\ Commun.\  {\bf 177} (2007) 399 [arXiv:hep-ph/0612134].

\bibitem{wmap}  
D.~N.~Spergel {\it et al.}  [WMAP Collaboration],
Astrophys.\ J.\ Suppl.\  {\bf 170} (2007) 377 [arXiv:astro-ph/0603452].


\end{thebibliography}
\end{document}